\documentclass[twocolumn,showpacs,preprintnumbers,amsmath,amssymb]{revtex4}

\begin{document}


\title[Non-Markovian Stochastic Processes]{Non-Markovian Stochastic
Processes\\ and the Wave-like Properties of Matter}

\author{Mario J. Pinheiro}
\email{mpinheiro@ist.utl.pt} \affiliation{Department of Physics
and Centro de F\'{i}sica de Plasmas, Instituto Superior
T\'{e}cnico, Av. Rovisco Pais, \& 1049-001 Lisboa, Portugal}

\homepage{http://alfa.ist.utl.pt/~pinheiro}

\thanks{}

\date{\today}

\begin{abstract}
A non-markovian stochastic model is shown to lead to a universal
relationship between particle's energy, driven frequency and a
frequency of interaction with the medium. It is briefly discussed
the possible relevance of this general structure to various
phenomena in the context of the formation of patterns in granular
media, computation in a Brownian-type computer and
the Haisch-Rueda-Puthoff inertial mass theory.
\end{abstract}


\maketitle

\bibliographystyle{apsrev}

\section{Introduction}

The study of physical systems with non-markovian statistical
properties has provided a natural basis for the understanding of
the role played by memory effects in such different fields as
anomalous transport in turbulent plasmas ~\cite{Ballescu95},
Brownian motion of macroparticles in complex fluids
~\cite{Amiblard96}, in the vortex solid phase of twinned
YBa$_2$Cu$_3$O$_7$ single crystals~\cite{Bekeris00}. Recently,
experimental evidence were reported of the quantum jumps of
particles in the Earth's gravitational field~\cite{Nature1} giving
a strong evidence that jumping process is quite ubiquitous in
natural processes. Thus, it becomes interesting to inquire: how
the medium interaction, perturbing the free motion of a particle,
leaves its own signature?

It is the purpose of this paper to provides the main lines of a
derivation of what we believe to be a nontrivial property of the
jumping process particles undergo, whenever they move in a
perturbing medium. The idea is to put a particle moving straight
in a discrete planar geometry, jumping from one site to another
and, in the meanwhile, subject to a random interaction.

It is found that a universal and structurally very simple expression of
the particle's energy prevails, intrinsically linked to the
generation of such properties as the mass of a particle and the
build-up of regular geometrical patterns.

This paper is organized as follows: In Section 2 we introduce, for
completeness, the basic features of a non-markovian stochastic
model. It is then shown that the classical particle's energy must
be proportional to the squared driven frequency over a frequency
of dissipation in the medium; in particular, whenever the
particle's dynamics is described by a planar wave, the de Broglie
relationship is retrieved. In Section 3, the evidence of the
referred general structure of the particle's energy is intuitively
discussed in several branches of physics. We believe that the
relationship between driving frequency of a given process and its
corresponding frequency of interaction with a medium provides a
method for a general approach to dissipative systems and is able
to predict a class of new relations.

\section{Infinite Memory Model}

In a non-markovian model the prediction about the next link
($x_{n+1}$) are defined in terms of mutually dependent random
variables in the chain ($x_1$, $x_2$,...,$x_n$). Consider a
particle jumping from one site to another in Euclidean space. We
will address here the much simple situation of a deterministic
jump process along a given direction. The jumping sites are
assumed to be equidistantly distributed along the axis. Now, add
to this jumping process an oscillatory motion due to interaction
with a medium and characterized by stochasticity. The frequency of
oscillation around an equilibrium position between two jumps is
denoted by $\nu$ and $\beta$ is the probability that each
oscillation in the past has to trigger a new oscillation in the
present. The simplicity of the described geometry is to same
extent well justified by the recent experiments done by
Nesvizhevsky and collaborators~\cite{Nature1}. Ultracold neutrons
in a vertical fall subject to a constant acceleration due to
gravity, were shown do not move continuously, but rather jump from
one height to another in quantum leaps.

Let $Q_m[q(t)]$ be the probability that one oscillation from the
$M=m_0+...+m_{q-1}$ which occurred in the past generates $m$
oscillations at the $qth$ step. Since we assume $\beta$ is
constant, this is an infinite memory model, meaning that an
oscillation which has occurred long time ago produces the same
effect as an oscillation which has occurred in the near past. Lets
introduce the probability density, $Q_n(t)dt$, that the $nth$
oscillation takes place in the interval of time $(t,t+dt)$ at
q$th$ step. Then
\begin{equation}\label{}
Q_{n+1}[q(t)]=\int_0^{q(t)} Q_{n}[q(t)] p_0[(q(t) - q(t')] d
q(t'),
\end{equation}
where $p_0(t-t')$ is the probability per unit time that the
$(n+1)st$ oscillation takes place in the time interval $(t,t+dt)$
given that the $nth$ oscillation took place at t'. Due to the
hidden interactions the particle undergo with the medium, we treat
the time of an oscillation as a random variable following a
Poisson distribution
\begin{equation}\label{}
p_0(t-t') = \left \{ \begin{array}{ll}
  0                        & \mbox{,if $(t-t') < \tau $} \\
  \nu d t \exp[-\nu(t-t')] & \mbox{,otherwise}.
\end{array}
\right.
\end{equation}
Here, $\nu$ is the frequency of an oscillation and $\tau$ is the
"dead" time. Designing by $\chi_n(s)$ and $\pi_0(s)$ the Laplace
transforms of $Q_n(t)$ and $p_0(t)$, resp., the convolution
theorem gives
\begin{equation}\label{}
\chi_{n+1}(s) = \chi_n(s)\pi_0(s).
\end{equation}
From this expression we obtain the recursive relation
\begin{equation}\label{}
\chi_n(s)=[\pi_0(s)]^{n-1} \chi_1(s).
\end{equation}
The evaluation of the transforms $\pi_0(s)$ and $\chi_1(s)$ gives
\begin{equation}\label{}
\pi_0(s)=\frac{\nu \exp(-(\nu+s)\tau)}{\nu + s},
\end{equation}
\begin{equation}\label{}
\chi_1(s)=\frac{\nu}{\nu +s},
\end{equation}
and
\begin{equation}\label{}
\chi_n(s)=\nu^n \frac{\exp(-(n-1)(\nu+s]\tau}{(\nu + s)^n}.
\end{equation}
The inverse transform calculated using the Laplace inverse
theorem, gives the probability for the occurrence of $m$
oscillations at time $t$:
\begin{equation}\label{}
Q_n(t)= \left\{ \begin{array}{ll} \nu
\frac{\{\nu[t-(n-1)\tau]\}^{n-1} \exp(-\nu t)}{(n-1)!} &
\mbox{,$t>(n-1)\tau$}\\
  0 & \mbox{,$t<(n-1)\tau$}.
\end{array}
\right.
\end{equation}
To simplify, we shall put $\tau=0$ and the probability density
that the $nth$ oscillation takes place in the interval of time
$(t,t+dt)$ reads
\begin{equation}\label{}
Q_n(t)d t=\frac{\nu(\nu t)^{n-1}}{(n-1)!}\exp(-\nu t) d t.
\end{equation}
It follows the probability density of occurrence of q-jumps at
time $t$ given by
\begin{equation}\label{}
\Psi_q(t) d t=\sum_{n=1}^{\infty} \xi_q(n) Q_n(t) d t,
\end{equation}
or, in complete form,
\begin{equation}\label{}
\Psi_q(t) d t = \sum_{n=1}^{\infty} \xi_q(n) \frac{\nu(\nu
t)^{n-1}}{(n-1)!} \exp(-\nu t) d t.
\end{equation}
Here, $\xi_q (n)$ is the probability to occur $n$ oscillations at
$qth$ jump. To evaluate $\xi_q(n)$ we first define $g_M(m_q)$, the
probability that M previous oscillations generate $m_q$
oscillations at $qth$ step~\cite{vlad}. The Bose-Einstein
distribution is favored since many oscillations can pertain to the
same step:
\begin{equation}\label{BED}
g_M(m_q) = \frac{(M + m_q - 1)!}{m_q!(M-1)!} \beta^{m_q} (1-\beta)^{M}.
\end{equation}

Introducing the conditional probability
$\varphi_q(m_q|m_{q-1},...,m_0)$ that at $qth$ step there are
$m_q$ oscillations provided that at the previous steps
$m_{q-1},...,m_0$ oscillations have occurred, subject to the
normalization condition
\begin{equation}\label{norcond}
\sum_{m_q} \varphi_q(m_q|m_{q-1},...m,m_0)=1,
\end{equation}
then, it can be shown ~\cite{vlad} that
\begin{displaymath}
\xi_q (n) = \sum_m \xi_0(m) (1-\beta)^{qm} [1-(1-\beta)^q]^{n-m}
\end{displaymath}
\begin{equation}\label{}
\frac{(n-1)!}{(m-1)! (n-m)!}.
\end{equation}
Hence, the probability density of occurrence of q-jumps is written
in the form
\begin{equation}\label{wp}
\Psi_q(t) d t = \frac{\alpha}{\nu} \exp(-\alpha t) \sum_m \xi_0(m)
\frac{(\alpha t)^{m-1}}{(m-1)!} d t,
\end{equation}
where we put $\alpha(q)\equiv (1-\beta)^q \nu$. It must be assumed
we know $\xi_0(m)$, that is the probability to occur $m$
oscillations from $t=0$ up to the first jump.

With the assumption of a Poisson distribution for $\xi_0(m)$, the
summation gives
\begin{equation}\label{gauss}
\sum_{m=0}^{\infty} \xi_0(m) \frac{(\alpha
t)^{m-1}}{(m-1)!}=\frac{1}{\sqrt{\lambda \alpha t}}
I_1(\sqrt{\lambda \alpha t}),
\end{equation}
where $I_1(x)$ is the first class modified Bessel function of
order 1. Hence, the final result for the probability of occurrence
of q-jumps between $t$ and $t+dt$ is given by
\begin{equation}\label{psi1}
\Psi_q(t) d t = \sqrt{\frac{\alpha}{\lambda \nu^2 t}} \exp(-\alpha
t) I_1(\sqrt{\lambda \alpha t}) dt.
\end{equation}
Eq.~\ref{psi1} is characterized by a temporal argument and, in
particular, for a sufficient number of steps, the limit
$\sqrt{\lambda x} \to 0$ is satisfied and Eq.~\ref{psi1} reduces
to
\begin{equation}\label{prob2}
\Psi_q (t) d t \approx \frac{\alpha}{2} \exp[-\alpha t] d t.
\end{equation}
We have in view a deterministic particle' system evolving
according to a local mapping in a space of equidistant sites. This
idealization lies in the Ehrenfest's equation describing the
quantum mechanical mean values of position, and thus avoids the
solution of a much more complex problem~\cite{vlad}. Hence, we can
rewrite the above equation in the form
\begin{equation}\label{prob3}
\Psi(x,t) dt \approx \frac{\alpha}{2} \exp [-\alpha t] dt.
\end{equation}
In terms of basic assumption of wave mechanics, the probabilities
are quadratic forms of $\psi$ functions,
$\Psi_q(t)=|\psi_q(t)|^2$, with $\psi$ the associated "wave".
Therefore, we can seek a representation of the transport process
in terms of wave function. By expanding the temporal argument
present in the exponential function in Eq.~\ref{prob3} and
retaining only terms of magnitude $\beta^2$ (higher order terms
are less important and are harder to give them a physical
meaning), we get
\begin{equation}\label{dev}
(1- \beta)^q \nu t \approx \nu t  - \frac{\beta q \Lambda
\nu}{\Lambda} t + \frac{\hbar}{2} \frac{(\beta q \nu)}{\hbar \nu}
t + \mathcal{O}(\beta^3).
\end{equation}

The expansion allowed the identification of some mechanical
properties of the particle in the medium. In fact, in analogy with
a transversal wave in a vibrant string, we can define a group
velocity $V \equiv \frac{\nu l}{2 \pi n}$, with $n=1,2,3,...$
(assuming a non-dispersive medium) as the instantaneous wave
velocity ($l$ is the average distance between two jumps) and $\nu$
as the number of cycles per second loosed on a given space
position ($\nu$ is an angular frequency), both quantities as seen
by an observer at rest in the lattice. As a first approximation
the hyperbolic partial differential equation is a good
representation of almost all electrodynamics, yielding plane wave
solutions in the complex representation $\Psi \sim \exp
\imath(\omega t-\overrightarrow{K}.\overrightarrow{r})$. According
to this representation we are lead to identify the third term on
the right-hand side of Eq.~\ref{dev} with the energy carried by
the particle:
\begin{equation}\label{kla}
E = \frac{\hbar}{2 \nu} \left( \beta q \nu \right)^2.
\end{equation}
The probability that each oscillation in the past has to trigger a
new oscillation in the present, $\beta$, can be defined by means
of the relativistic formula for energy, yielding $\beta = \frac{f
\Lambda}{c}$~\cite{Note1}, where $\Lambda \equiv \frac{l}{q}$
represents a distance over number of jumps, that is, it is an
average distance per step; $f$ represents the role of the driven
frequency since through $\beta$ the wave propagation is implicitly
included. But $E=\overline{m}c^2$, inasmuch as Einstein's
relationship holds true. Finally, putting $\omega = 2 \pi f$, it
is interesting that we get the following universal and
structurally very simple expression for the mass
\begin{equation}\label{Broglie2}
\overline{m} = \frac{\hbar \omega^2}{\nu c^2} n^2.
\end{equation}
The integer $n$ allow the possible existence of sub-harmonics in
the energy spectrum. So far, the above equation belongs to a set
of structural relationship between some physical concepts: energy,
driven frequency $\omega$ and dissipative frequency $\nu$. In
fact, $\nu$ embodies the degree of interaction with the medium,
but $\beta$ is intrinsically related to the structure of the
medium itself. Hence, the non-markovian character of the
stochastic process is related to the nature of the medium rather
than the past history of the particle.

The medium perturbation is characterized by $\nu$ shaping the
coefficient $\overline{m}$ in a slightly different pattern than
the well known quantum mechanical expression for a photon packet
($E=\hbar \omega$). Note that Eq.~\ref{Broglie2} is consistent
with the De Broglie relation for free particles (planar waves),
since then $\omega=\nu$ (resonance condition). Otherwise, the
dissipation in the medium must be taken into account to verify a
different relationship for the particle´s energy, that is,
Eq.~\ref{Broglie2}. We would like to point out that, in as much as
a quantitative correlation is match between a source system
(providing the periodic force which delivers energy to a given
process) and an interactive medium, our Eq.~\ref{Broglie2} has a
possible connection with the generalized Nyquist
relation~\cite{Callen}. Also, it is worth to remark that our main
result is not merely a bookkeeping of phenomena, but it traduces a
balance between two different physical mechanisms related to the
driving force and interactive process.

\section{Application in various branches of physics}

Having presented the mainframe of the infinite memory model, we
give now some illustrations upon our point of view that it is
possible to achieve an understanding of some physical phenomena
in, possibly, a wide range. In the examples given below the
general structure described by Eq.~\ref{Broglie2} seems to govern
geometrical patterns in granular media, inertial effects in
stochastic electrodynamics and minimal energy expenditure in
computational thermodynamics.

All phenomena hereby referred share the same imprint of a
stochastic interaction with a medium, building-up a given physical
structure.

\subsection{Spheres packing and gravitational surface waves}

Granular matter consists of macroscopic particles of different
size, shape and surface properties and those characteristics lead
to specific packing behavior. Particle clustering results from an
energy loss associated with particle-particle interactions. This
interesting and fascinating behavior can be described by the
infinite memory model and the constitutive equation embodied in
~\ref{Broglie2}. To probe it, we start by rearranging
Eq.~\ref{Broglie2} in the form
\begin{equation}\label{}
\lambda =\frac{\overline{m}c^2 v }{h f^2}= \frac{\frac{E
v}{h}}{f^2}=\frac{g_{eff}}{f^2}.
\end{equation}
Here, we defined $v=\frac{\lambda \nu}{2 \pi}$ and $g_{eff} =
\frac{E v}{h}$ as an effective acceleration. $\lambda$ is the
wavelength associated to the geometrical pattern observed.
Detailed experiments on spheres packing of diameter $d$ in 2 and 3
dimensions have shown~\cite{Duran} the display of repetitive
geometric patterns, similar to the instabilities reported by
Faraday.

Actually, it was observed a dependence of the wavelength of those
geometrical patterns and the frequency of excitation $f$ imposed
vertically on a thin layer of granular matter. Both are related
through the equation
\begin{equation}\label{}
\lambda = \lambda_{min} + \frac{g_{eff}}{f^2},
\end{equation}
where $\lambda_{min}$ represents a threshold near $11 d$, with $d$
denoting the particles diameter. Incidentally, gravitational waves
in the surface of a fluid have the same dependency~\cite{Landau}.

The two main mechanisms governing the phenomena are the direct
excitation of surface waves and a mechanism of successive
bifurcations resulting from the excitations due to the vibrations
of granular matter. In fact, memory-effects have been
experimentally shown to occur in granular
materials~\cite{Jaeger2000}.

\subsection{Computation in a Brownian-type computer}

The work on classical, reversible computation has laid the
foundation for the development of quantum mechanical computers.

In 1961 Landauer analyzed the physical limitations on computation
due to dissipative processes~\cite{Landauer61}. He showed that
energy loss could be made as small as you want, provided the
device works infinitesimally slowly. In fact, it is shown that the
first condition for any deterministic device to be reversible is
that its input and output be uniquely recoverable from each other
- this is the condition of logical reversibility. If, besides this
condition, a device can actually run backwards, then it is called
physically reversible and it dissipates no heat, according to the
second law of thermodynamics.

An example of reversible computing involves a copying machine, for
example. Feynman derived a formula estimating the amount of free
energy it takes to realize a computation in a given interval of
time~\cite{Feycomp}. Envisioning a computer designed to run by a
diffusion process, characterized by a slightly higher probability
to run forward than backwards, Feynman proposed the relationship
to hold
\begin{equation}\label{Fey1}
E = k_BT \frac{t_m}{t_a}.
\end{equation}
Here, $E$ is the energy loss per step, $k_B T$ is thermal energy,
$t_m$ and $t_a$ are, resp., the minimum time taken per step and
the time per step actually taken. It is easily seen that the $k_B
T=\hbar \omega$ holds true since thermal energy is the driven
process and $\omega = \frac{2 \pi}{t_a}$ and $\nu = \frac{2
\pi}{t_m}$. The forward transition rate to a new configuration of
available states (say from $\{n_i\}$ to $\{n_j\} > \{n_i \}$) span
in a time scale which has a non-null correlation factor and thus
generating a memory effect. Although, as stressed by Feynman,
Eq.~\ref{Fey1} is only approximative, we view in it a particular
manifestation of Eq.~\ref{Broglie2}. In fact, Eq.~\ref{Fey1}
represents the minimum energy that must be expended per
computational step in a given process. We have actually a new deep
insight to our Eq.~\ref{Broglie2}: it results from the best match
between energy cost versus speed. It is a by-product of minimum
principles.

\subsection{Haisch-Rueda-Puthoff inertial mass theory}

Based on stochastic electrodynamics, Haisch, Rueda and
Puthoff~\cite{Puthoff} put in evidence the relationship between
the zero-point field (ZPF) and inertia. ZPF is uniform and
isotropic in inertial frames, while showing asymmetries when
viewed in accelerated frames. Applying a technique developed
formerly by Einstein and Hopf~\cite{Hopf} and which is at the
foundation of stochastic electrodynamics, the charged particles
constituent of matter (partons or quarks) were driven to oscillate
at velocity $\mathbf{v}_{osc}$ by the electric component of the
ZPF, $\mathbf{E}^{ZP}$, thereby accelerating in a direction
perpendicular to the oscillations induced by the ZPF. The action
of the magnetic component of the ZPF generate a Lorentz force
whose average value is given by
\begin{equation}\label{}
\mathbf{F}_L = < \left(\frac{q}{c} \right) [ \mathbf{v}_{osc}
\times \mathbf{B}^{ZP} ] > = -\frac{\Gamma \hbar \omega_c^2}{2 \pi
c^2} \mathbf{a}.
\end{equation}
They interpreted this result as an account for inertia and the
inertial mass, $m_i$, is shown to be a function of the damping
constant for the oscillations and $\omega_c$, the characteristic
frequency of particle-ZPF interactions:
\begin{equation}\label{put1}
m_i = \frac{\Gamma \hbar \omega_c^2}{2 \pi c^2}.
\end{equation}
In the above expression, $\Gamma=\frac{q^2}{6 \pi \epsilon_0 m_0
c^2}$ is the Abraham-Lorentz radiation damping constant appearing
on the nonrelativistic equation of motion for a particle of mass
$m_0$ and charge $q$ when submitted to the zero-point radiation
electric field; $\omega_c= \sqrt{\frac{\pi c^5}{\hbar G}}$ is the
effective Planck cutoff frequency of the vacuum
zero-point-fluctuation spectrum ~\cite{Puthoff89}. This idea is
rooted in a former publication by Sakharov~\cite{Sakharov}
envisioning gravitation as resulting from a small disturbance of
the metrical elasticity of space.

Comparing Eq.~\ref{put1} with Eq.~\ref{Broglie2}, it is clear that
those expressions have the same structure. Interestingly, Puthoff
and collaborators conjectured that their interpretation of mass as
resulting from a resonance with the ZPF leads directly to the de
Broglie relation ~\cite{HR-kluwer}. In their interpretation,
inertia is a kind of electromagnetic drag that affects charged
particles undergoing acceleration through the electromagnetic
zero-point field. In fact, according to the de Broglie
perspective, the inertial mass of a particle is the vibrational
energy divided by c$^2$ of a localized oscillating field. It is
not an intrinsic property but instead a measure of the degree of
coupling with a localized field (already, in his own view, of
electromagnetic origin).

However, it will be noted that our analysis gives evidence of a
more complex structure although, when the resonant condition is
verified, it collapses to the well-known the Broglie relation.

\section{Conclusion}

Exploring the underlying transport mechanism of a test particle
with infinite memory induces us to attribute a universal and
structurally simple property to the particle's energy, embodied in
Eq.~\ref{Broglie2}. In our interpretation, in a perturbative
medium the particle's energy results from a balance between the
driven frequency $\omega$ and a frequency of interaction in the
medium, $\nu$. In the particular case of planar waves this result
is consistent with de Broglie wavelength relationship.

Our approach incorporates the fundamental properties of dynamics
and how deeply rooted in natural phenomena is this general
structure, it is illustrated in various objects and different kind
of fields.

\begin{acknowledgments}
I wish to acknowledge Prof. Paulo Sá, from the University of
Porto, for the critical reading of the manuscript.
\end{acknowledgments}

\bibliographystyle{amsplain}

\end{document}